\documentclass[journal]{IEEEtran}

\usepackage[utf8]{inputenc}
\usepackage{amsmath,amsfonts}
\usepackage{cite}
\usepackage{multicol}
\usepackage{graphicx}
\usepackage{array}
\usepackage{multirow}
\usepackage{graphicx}
\usepackage{enumitem}
\usepackage{color}
\usepackage{nomencl}
\usepackage{etoolbox}
\usepackage{amssymb}
\usepackage{comment}
\usepackage{mathtools}
\usepackage{bbold}
\usepackage{yfonts}
\usepackage{soul}
\usepackage{hyperref}
\usepackage{dsfont}
\usepackage[acronym]{glossaries}
\usepackage{nomencl}
\usepackage{subfiles}
\usepackage[utf8]{inputenc}
\usepackage{mathtools}
\usepackage{amsmath}
\usepackage{tikz}
\usepackage{algorithm}
\usepackage{algpseudocode}
\usepackage{etoolbox}
\usepackage{fnpct}
\ifCLASSOPTIONcompsoc \usepackage[caption=false,font=normalsize,labelfon
t=sf,textfont=sf]{subfig}
\else
\usepackage[caption=false,font=footnotesize]{subfig}
\fi
\usepackage{tikz}

\ifCLASSINFOpdf
\else
\fi

\begin{document}

\title{Fair Cost Allocation in Energy Communities:\\ A DLMP-based Bilevel Optimization with a\\ Shapley Value Approach}

\author{
\IEEEauthorblockN{Hyeongon Park\thanks{School of Systems Management and Safety Engineering, Pukyong National University, Busan, 48516, South Korea (e-mail:\ hyeongon@pknu.ac.kr)}, 
Kyuhyeong Kwag\thanks{Market Innovation Department, Korea Power Exchange, Jeollanam-do, 58322, South Korea (e-mail: k.kwag@kpx.or.kr)}, 
Daniel K. Molzahn\thanks{School of Electrical and Computer Engineering, Georgia Institute of Technology, Atlanta, GA, 30313, USA (e-mail:\ molzahn@gatech.edu)}, and 
Rahul K. Gupta\thanks{School of Electrical Engineering and Computer Science, Washington State University, Pullman, WA, 99163, USA (e-mail:\ rahul.k.gupta@wsu.edu)}}
}

\makeatletter
\patchcmd{\@maketitle}
  {\addvspace{0.5\baselineskip}\egroup}
  {\addvspace{-2\baselineskip}\egroup}
  {}
  {}
\makeatother
\maketitle
\begin{abstract}
Energy communities (ECs) are emerging as a promising decentralized model for managing cooperative distributed energy resources (DERs). As these communities expand and their operations become increasingly integrated into the grid, ensuring fairness in allocating operating costs among participants becomes a challenge. In distribution networks, DER operations at the community level can influence Distribution Locational Marginal Prices (DLMPs), which in turn affect system's operation cost. This interdependence between local decisions and system-level pricing introduces new challenges for fair and transparent cost allocation. Despite growing interest in fairness-aware methods, most methods do not account for the impact of DLMPs. To fill this gap, we propose a bilevel optimization model in which a Community Energy Aggregator (CEA) schedules DERs across multiple ECs while a Distribution System Operator (DSO) determines DLMPs through network-constrained dispatch. Leveraging the Karush-Kuhn-Tucker (KKT) conditions and strong duality, the bilevel model is reformulated into a tractable single-level problem. We achieve fairness in the cost allocation by applying the Shapley value to quantify each community’s marginal contribution to system-wide cost savings. The effectiveness of the proposed method is validated through simulations on several benchmark distribution systems.
\end{abstract}


\IEEEpeerreviewmaketitle
\section*{Nomenclature}

\subsubsection*{Sets}
\begin{IEEEdescription}[\IEEEusemathlabelsep\IEEEsetlabelwidth{$u^{\text{ch}}_{l,t}, u^{\text{dis}}_{l,t}$}]
\item[$\mathcal{N}$] Set of all distribution buses.
\item[$\mathcal{N}_\text{c}$] Set of buses where communities are located.
\item[$\mathcal{N}_s$] Set of slack buses (single bus at index $0$).
\item[$\mathcal{L}$] Set of distribution lines $(l,k)$ in the network.
\item[$\mathcal{T}$] Set of time periods in the scheduling horizon.
\end{IEEEdescription}

\vspace{0.5em}
\subsubsection*{Parameters}
\begin{IEEEdescription}[\IEEEusemathlabelsep\IEEEsetlabelwidth{$u^{\text{ch}}_{l,t}, u^{\text{dis}}_{l,t}$}]
\item[$\hat{P}^{\text{pv}}_{l,t}$] Forecasted available PV generation at bus~$l$ and time~$t$.
\item[$S^{\text{pv}}_{l,\text{max}}$] Apparent power rating of PV inverter at bus~$l$.
\item[$\text{PF}_{\text{min}}$] Minimum allowable power factor of PV inverter.
\item[$\zeta^{\text{pv}}$] Reactive power coefficient derived from $\text{PF}_{\text{min}}$.
\item[$\overline{P}^{\text{ch}}_l$] Charging power capacity of BESS at bus~$l$.
\item[$\overline{P}^{\text{dis}}_l$] Discharging power capacity of BESS at bus~$l$.
\item[$\underline{E}_l$, $\overline{E}_l$] Min, max energy limits of BESS at bus~$l$.
\item[$\eta^{\text{ch}}_l$, $\eta^{\text{dis}}_l$] Charging and discharging efficiency of BESS at bus~$l$.
\item[$\hat{P}^{\text{load}}_{l,t}, \hat{Q}^{\text{load}}_{l,t}$] ~Forecasted active and reactive load of community at bus~$l$ and time~$t$.
\item[$\overline{P}^{\text{flex}}_{l,t}$] Maximum flexible portion of controllable load at bus~$l$ and time~$t$.
\item[$\pi^{\text{flex}}$] Compensation rate for curtailed load.
\item[$r_{kl}$, $x_{kl}$] Resistance and reactance of line $(k,l)$.
\item[$z_{kl}$] Complex impedance of line $(k,l)$, i.e., $z_{kl} = r_{kl} + jx_{kl}$.
\item[$\pi^{\text{LMP}}_t$] Wholesale electricity price at time $t$.
\item[$c^g_l$] Marginal cost of local generator at bus $l$.
\item[$\underline{p}^g_l$, $\overline{p}^g_l$] Minimum and maximum active power output of generator at bus $l$.
\item[$\underline{q}^g_l$, $\overline{q}^g_l$] Minimum and maximum reactive power output of generator at bus $l$.
\item[$v^{\text{min}}$, $v^{\text{max}}$] Lower, upper limits of bus voltage magnitude.
\end{IEEEdescription}

\vspace{0.5em}
\subsubsection*{Decision Variables}
\begin{IEEEdescription}[\IEEEusemathlabelsep\IEEEsetlabelwidth{$u^{\text{ch}}_{l,t}, u^{\text{dis}}_{l,t}$}]
\item[$p^{\text{pv}}_{l,t}$, $q^{\text{pv}}_{l,t}$] Active and reactive power outputs of PV system at bus~$l$ and time~$t$.
\item[$p^{\text{ch}}_{l,t}$, $p^{\text{dis}}_{l,t}$] Charging and discharging power of BESS.
\item[$u^{\text{ch}}_{l,t}$, $u^{\text{dis}}_{l,t}$] Binary indicators for charging and discharging mode of BESS.
\item[$e_{l,t}$] State of Charge of BESS.
\item[$p^{\text{bat}}_{l,t}$] Net active power from BESS ($p^{\text{ch}}_{l,t} - p^{\text{dis}}_{l,t}$).
\item[$p^{\text{load}}_{l,t}$] Scheduled controllable load.
\item[$p^{\text{c,red}}_{l,t}$] Amount of curtailed controllable load.
\item[$p^c_{l,t}$, $q^c_{l,t}$] Net active and reactive power consumption of the community.
\item[$\lambda^p_{l,t}$, $\lambda^q_{l,t}$] DLMPs for active and reactive power at bus~$l$ and time~$t$.
\item[$p^g_{l,t}$, $q^g_{l,t}$] Active and reactive power outputs of the generator at bus $l$ and time $t$.
\item[$u_{l,t}$] Squared voltage magnitude at bus $l$ and time $t$.
\item[$P_{lk,t}$, $Q_{lk,t}$] ~Active and reactive power flows from bus $l$ to $k$ at time $t$.
\item[$S_{lk,t}$] Complex power flow from bus $l$ to $k$ at time $t$.
\item[$f_{lk,t}$] Current magnitude on line $(l,k)$ at time $t$.
\item[$p^{\text{inj}}_{l,t}$, $q^{\text{inj}}_{l,t}$] Net active and reactive power injection at bus $l$ at time $t$.
\end{IEEEdescription}

\vspace{0.5em}
\subsubsection*{Lagrange Multipliers}
\begin{IEEEdescription}[\IEEEusemathlabelsep\IEEEsetlabelwidth{$u^{\text{ch}}_{l,t}, u^{\text{dis}}_{l,t}$}]
\item[$\lambda^{\text{volt}}_{lk,t}$] Multiplier for voltage drop constraint on line $(l,k)$ at time $t$.
\item[$\underline{\lambda}^u_{l,t}$, $\overline{\lambda}^u_{l,t}$] Multipliers for lower and upper bounds on squared voltage magnitude at bus $l$.
\item[$\lambda^p_{l,t}$, $\lambda^q_{l,t}$] Multipliers for active and reactive power balance at bus $l$ and time $t$.
\item[$\underline{\lambda}^{gp}_{l,t}$, $\overline{\lambda}^{gp}_{l,t}$] Multipliers for lower and upper bounds on $p^g_{l,t}$.
\item[$\underline{\lambda}^{gq}_{l,t}$, $\overline{\lambda}^{gq}_{l,t}$] Multipliers for lower and upper bounds on $q^g_{l,t}$.
\end{IEEEdescription}

\section{Introduction}
The rapid proliferation of prosumers equipped with renewable energy resources (RES) and battery energy storage systems (BESS) has significantly increased interest in decentralized energy systems in recent years \cite{de2021smart}. One such example is the rise of energy communities (ECs) \cite{EU_directive2023},
which coordinate distributed energy resources (DERs) and flexible loads at the local level. ECs have emerged as a promising solution to promote sustainability, resilience, and local energy autonomy \cite{EU_directive2023}. While ECs offer various technical and economic benefits, such as improved self-consumption, peak shaving, and reduced grid dependency, their effective formation and operation pose several challenges. A critical issue lies in the fair allocation of collective costs and benefits arising from shared assets or coordinated operation within the distribution grid
\cite{cremers2023efficient}.

Recent studies on collaborative energy systems (e.g., ECs) have increasingly explored how to fairly allocate costs and benefits among participants. 
For example, \cite{malik2022priority} adopts a Nash bargaining–based priority mechanism for peer‑to‑peer energy trading in local communities, and a nucleolus analysis is used in \cite{churkin2021enhancing} to ensure coalition stability in transmission expansion planning models. The Shapley value \cite{shapley1953value} has also attracted attention as an approach for fair cost and benefit allocation.

This approach is designed to allocate the total cost (or benefit) based on each participant’s marginal contribution across all possible groupings and is widely regarded for satisfying key fairness principles such as efficiency, symmetry, and the null player property \cite{WINTER2002}. These axioms make the Shapley value a fair and transparent solution for cooperative cost sharing, particularly in settings where multiple parties share infrastructure or coordinate their operations.

Building on this idea, several studies have applied the Shapley value in the context of investment allocation for ECs. For instance, Rui et al.\cite{rui2024incentive} proposed a Shapley-based incentive scheme to allocate the cost of grid-enhancing technologies fairly among multiple participants. Similarly, Pedrero et al.\cite{pedrero2024fair} introduced a scalable Shapley value method to determine fair investment strategies for large energy communities, focusing on equitable distribution of infrastructure costs such as shared batteries or RES installations.

The Shapley value has also been incorporated into the design of market mechanisms. Xie and Chen\cite{xie2024real} employed an Aumann–Shapley pricing scheme to derive real-time bidding strategies for storage operators, linking emission allocation and marginal system cost. Vespermann et al.~\cite{vespermann2020access} explored market designs that enable “access economy” models for energy storage, allowing non-owners to benefit from storage through financial or physical rights. Both \cite{xie2024real} and \cite{vespermann2020access} highlight the importance of cooperative cost or value allocation in market settings, using Shapley-based approaches to ensure fairness in either emission pricing or shared storage access. 

Despite the growing interest in fairness-aware mechanisms, existing works do not account for the dynamic feedback between community-level decisions and distribution network-level price signals. If benefits are not fairly allocated to ECs, each individual EC could be incentivized to adjust its operation to reduce the price at their own location, even if such actions increase the prices at other ECs. This behavior would ultimately deviate from the overall optimum for the group of ECs, reducing collective benefits. Fair allocation is thus essential to keep ECs aligned and incentivized to collaborate rather than compete.

In future distribution systems, node-level electricity prices are proposed through Distribution Locational Marginal Prices (DLMPs) \cite{papavasiliou2017analysis}, which reflect the spatiotemporal value of energy and grid constraints. To date, DLMPs have been widely studied as pricing signals but have not been applied to fair cost allocation among ECs in the same network.
In practice, the operation of each EC can influence DLMPs through its net injections and location. Thus, an EC’s scheduling decision affects not only its own cost but also system-wide DLMPs and cost distribution. While several studies have proposed Shapley-based methods for fair operating cost allocation in ECs \cite{eichelbeck2024fair, hupez2020new, chics2017coalitional, yang2021optimal, cremers2023efficient}, none of them consider the DLMPs variations due to ECs' coordination. For example, \cite{eichelbeck2024fair} focuses on allocation under forecast uncertainty, and \cite{hupez2020new} proposes a cooperative cost allocation rule based on predefined sharing functions. Similarly, \cite{yang2021optimal, chics2017coalitional} adopt coalitional game theory but assume fixed price environments or simplified distribution models. 

Moreover, the high computational complexity of cooperative game theory allocation methods, such as the Shapley value and the Nucleolus, is a well-known barrier to their application in large-scale systems \cite{OBrien2016, Cremers2023, Han2021, Ruiz2007, Churkin2024, Molina2014, Hupez2021}. To address this, approximation methods based on statistical sampling have been proposed, such as the stratified sampling highlighted in \cite{OBrien2016}. Another stream of research suggests grouping or clustering players to reduce the problem's dimensionality. However, the criteria for grouping in these studies are typically based on players' operational and economic characteristics, such as consumption patterns \cite{Cremers2023}, their energy profiles \cite{Han2021}, or market impact \cite{Ruiz2007}. While these approaches reduce complexity, they do not directly account for the underlying physical characteristics of the power grid, which fundamentally determine the players' interactions. The potential of using physical location and parameters for clustering has been noted as a promising future direction but remains an open research question \cite{Churkin2024}.

In light of these gaps, this paper proposes a fair cost allocation framework for a Community Energy Aggregator (CEA) coordinating multiple DER-equipped ECs accounting for the DLMP-based pricing. The problem is formulated as a bilevel optimization problem accounting for the interaction between the CEA and the Distribution System Operator (DSO). The scheme determines the DLMPs, which are then used to determine fair cost allocation using the Shapley value. We achieve scalability for large-scale systems via a novel signature-based approximation scheme to compute the the Shapley value. The proposed framework can offer valuable policy insights into (i) fairness-aware community coordination, (ii) incentive-compatible market design, and (iii) mechanisms to encourage DER participation in distribution-level markets.

The key contributions of this work are listed below.
\begin{enumerate}
    \item We formulate a DLMP-based bilevel optimization model that captures the interaction between a CEA and the DSO, including operational constraints and DER coordination across multiple communities.
    \item We develop a tractable single-level reformulation of the bilevel model using the Karush-Kuhn-Tucker (KKT) conditions and strong duality, enabling numerical solution under realistic settings.
    \item We propose a Shapley value-based cost allocation method that quantifies each community’s marginal contribution under DLMP feedback, and we evaluate its performance through case studies on benchmark distribution networks.
   \item We propose a signature-based approximation method for efficient Shapley value computation by grouping ECs based on their physical and topological characteristics.
\end{enumerate}

The remainder of this paper is organized as follows. Section~\ref{sec:Framework} describes the operational policies of the CEA and the DSO, including the problem structure and mathematical formulation. Section~\ref{sec:allocation} presents the coalitional game-theoretic model and details the Shapley value method for fair cost allocation among communities. In Section~\ref{sec:numerical_validation}, case studies are provided to evaluate the proposed methods, and Section~\ref{sec:conclusion} summarizes the findings and concludes the paper.
    \vspace{-1.0em}
\begin{figure}[!htbp]
    \centering
    \includegraphics[width=1.0\linewidth]{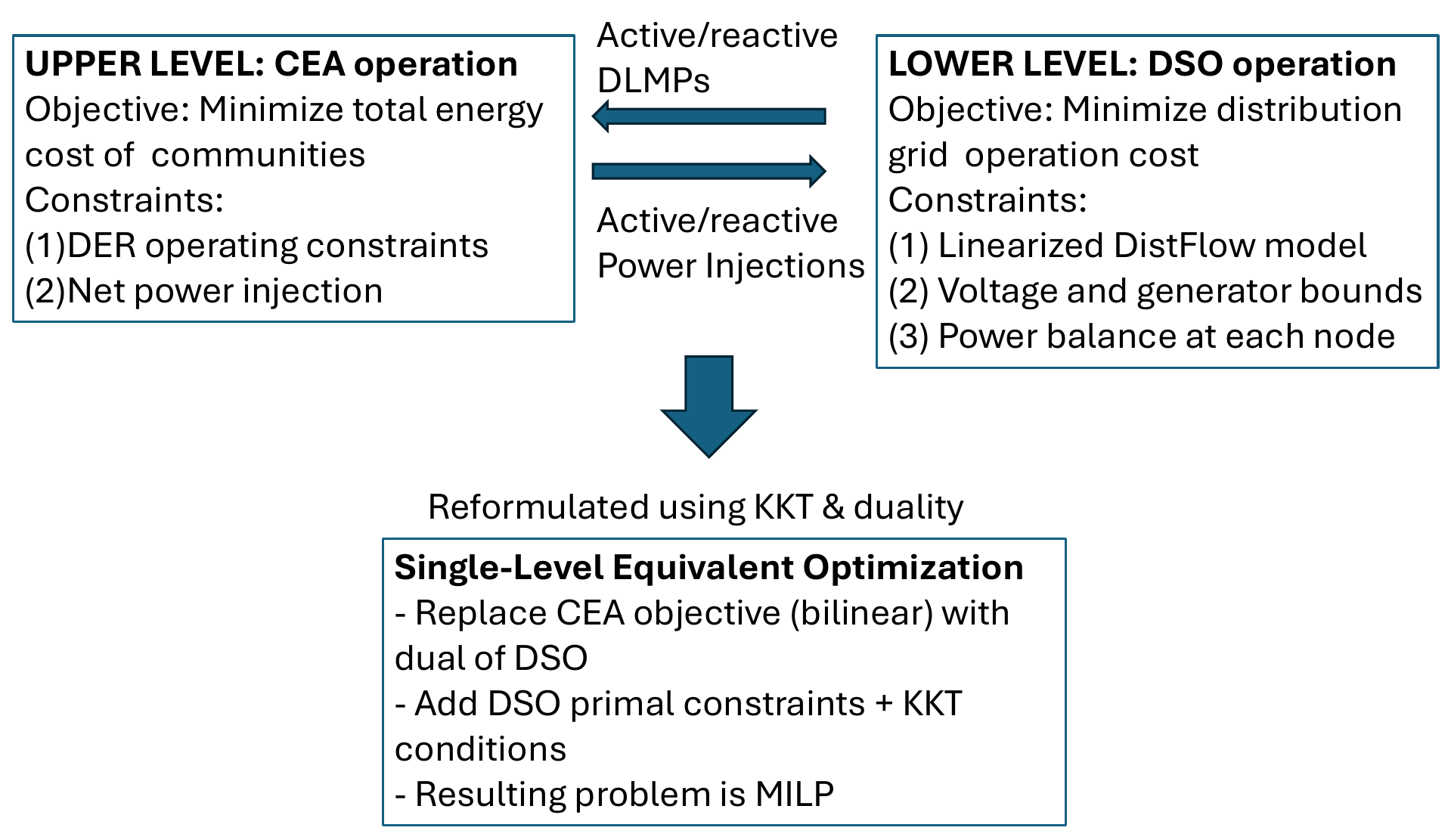}
    \vspace{-1.5em}
    \caption{Bilevel optimization framework between the CEA and the DSO, and its duality-based single-level reformulation.}
    \vspace{-1.5em}
    \label{fig:cea_dso_framework}
    
\end{figure}

\section{CEA \& DSO Scheduling Framework}
\label{sec:Framework}

\subsection{Description of Problem Structure}
\label{sec: Description_of_structure}
To enable coordination between local communities and the distribution network, this study models their interaction using a bilevel optimization framework, where the CEA and the DSO are structured as the leader and the follower, respectively. The overall structure of this interaction is illustrated in Fig.~\ref{fig:cea_dso_framework}. The upper-level problem represents the CEA's decision-making on community-level energy scheduling, while the lower-level problem reflects the DSO's response in maintaining system-wide balance and operational feasibility.

The CEA coordinates multiple ECs through centralized scheduling. These communities host various DERs that are directly controlled and optimized by the CEA. Specifically, photovoltaic (PV), BESS, and controllable loads are managed collectively to achieve the aggregate objectives of all communities. The individual contributions of each community can later be evaluated using various allocation methods.

The DSO is responsible for the reliable operation of the distribution network. It procures electricity from the transmission-distribution interface at wholesale market prices, supplies power to non-community consumers, and controls local generators while respecting distribution network constraints. As part of this process, the DSO must account for each community’s nodal power generation or consumption scheduled by its CEA.

The price signals used in CEA scheduling are DLMPs derived from the DSO’s optimization problem. These DLMPs are the dual variables associated with active and reactive power balance constraints at each bus. Consequently, each CEA's cost is affected by the locational and temporal conditions of the distribution grid. At the same time, the CEAs’ scheduling decisions influence the nodal power injections affecting the DSO’s power balance and the resulting DLMPs. This interdependency results in a coupled bilevel optimization problem. Similar bilevel structures in distribution systems, where price signals are determined by one entity and energy volumes are scheduled in response, have been studied in \cite{jangid2024distribution,park2022bi}.

\subsection{CEA Model}
\label{sec:CEA}

The CEA centrally coordinates DERs — including PV, BESS, and controllable loads — across multiple participating ECs. Each EC is located at a specific bus in the network, and the CEA optimizes their aggregated operation to minimize the total system cost under DLMP-based price signals.

The objective of the CEA is to minimize the total cost of active and reactive power consumption, along with the incentive payments for flexible load reductions. The corresponding objective function is given by

\begin{equation}
f^{\text{CEA}} = \sum_{t \in \mathcal{T}} \sum_{l \in \mathcal{N}_\text{c}} \left( 
\lambda^p_{l,t} \cdot p^c_{l,t} + 
\lambda^q_{l,t} \cdot q^c_{l,t} + 
\pi^{\text{flex}} \cdot p^{\text{c,red}}_{l,t}
\right)
\label{eq:cea_obj}
\end{equation}

Here, $\lambda^p_{l,t}$ and $\lambda^q_{l,t}$ represent DLMPs for active and reactive power and are treated as input parameters in the CEA problem. They are, however, determined as dual variables associated with power balance constraints in the DSO model described later in Section~\ref{sec:DSO}. It is important to note that the CEA only considers community buses in its objective function and excludes the operation of other non-community buses. 

The CEA scheduling problem includes constraints on DER operations across all community buses \( l \in \mathcal{N}_c \) and time periods \( t \in \mathcal{T} \).  These constraints consist of (i) PV generation limits, (ii) BESS operating rules, (iii) controllable load bounds, and (iv) the resulting net power consumptions submitted to the DSO. The complete set of constraints is outlined as follows:
\begin{subequations}  \label{eq:cea_constraints}
\begin{align}
    & {0} \leq {p}_{l,t}^{\text{pv}}  \leq \hat{{P}}_{l, t}^{\text{pv}} \label{eq:pv_model_1} \\
    & ({p}_{l, t}^{\text{pv}})^2 + ({q}_{l, t}^{\text{pv}})^2 \leq  ({S}^{\text{pv}}_{l,\text{max}})^2  \label{eq:pv_model_4} \\
        &  {q}_{l, t}^{\text{pv}} \leq {p}_{l, t}^{\text{pv}}\zeta^{\text{pv}}  \label{eq:pf1_ch6}\\
    &  -{q}_{l, t}^{\text{pv}}  \leq {p}_{l, t}^{\text{pv}}\zeta^{\text{pv}}  \label{eq:pf2_ch6}\\
    & 0 \leq p^{\text{ch}}_{l,t} \leq \overline{P}^{\text{ch}}_l \cdot u^{\text{ch}}_{l,t}  \label{eq:charge_limit} \\
    & 0 \leq p^{\text{dis}}_{l,t} \leq \overline{P}^{\text{dis}}_l \cdot u^{\text{dis}}_{l,t} \label{eq:discharge_limit} \\
    & u^{\text{ch}}_{l,t} + u^{\text{dis}}_{l,t} \leq 1 \label{eq:mutual_exclusivity} \\
    & e_{l,t+1} = e_{l,t} + \eta^{\text{ch}}_l p^{\text{ch}}_{l,t} \Delta t - \frac{1}{\eta^{\text{dis}}_l} p^{\text{dis}}_{l,t} \Delta t \label{eq:energy_balance} \\
    & \underline{E}_l \leq e_{l,t} \leq \overline{E}_l \label{eq:soc_bounds} \\
    & p^{\text{bat}}_{l,t} = p^{\text{dis}}_{l,t} - p^{\text{ch}}_{l,t} \label{eq:net_battery_power} \\
    & 0 \leq p^{\text{load}}_{l,t} \leq \hat{P}^{\text{load}}_{l,t} \label{eq:load_upper}
    \end{align}
    \begin{align}
    & p^{\text{c,red}}_{l,t} = \hat{P}^{\text{load}}_{l,t} - p^{\text{load}}_{l,t} \label{eq:load_reduction_def} \\
    & 0 \leq p^{\text{c,red}}_{l,t} \leq \overline{P}^{\text{flex}}_{l,t}  \label{eq:load_reduction_limit} \\
&p^c_{l,t} = p^{\text{load}}_{l,t} - p^{\text{pv}}_{l,t} - p^{\text{bat}}_{l,t} 
\label{eq:net_power_consumption} \\
&q^c_{l,t} = \hat{Q}^{\text{load}}_{l,t} - q^{\text{pv}}_{l,t}
\label{eq:net_reactive_power_consumption}
\end{align}
\end{subequations}

Constraints~\eqref{eq:pv_model_1}--\eqref{eq:pv_model_4} specify the PV operating limits i.e., real power availability based on solar forecasts, inverter capacity constraints on apparent power, and minimum power factor constraints reflecting grid code requirements. 

Constraints~\eqref{eq:charge_limit}--\eqref{eq:net_battery_power} are imposed on BESS units. Constraints~\eqref{eq:charge_limit} and~\eqref{eq:discharge_limit} set the maximum charging and discharging power levels, while~\eqref{eq:mutual_exclusivity} restricts simultaneous charging and discharging through binary control logic. The state of charge is updated in~\eqref{eq:energy_balance} considering BESS efficiency, and it is bounded by~\eqref{eq:soc_bounds}. Finally, ~\eqref{eq:net_battery_power} defines the net battery power as the difference between discharging and charging power.

Constraints~\eqref{eq:load_upper}--\eqref{eq:load_reduction_limit} describe the operation of controllable loads. Constraint~\eqref{eq:load_upper} sets the upper bound of the scheduled load based on forecasted demand, while~\eqref{eq:load_reduction_def} defines the curtailed portion as the difference from the forecast which is limited by \eqref{eq:load_reduction_limit}, defined by the user.

Constraints~\eqref{eq:net_power_consumption} and~\eqref{eq:net_reactive_power_consumption} compute the net active and reactive power consumptions at each community bus. These are obtained by subtracting local DER contributions, namely PV generation and battery discharge from the total load. Here, we assume that BESS does not provide any reactive power regulation. The resulting values are passed to the DSO as inputs to grid-level optimization and DLMP calculation.
\subsection{DSO Model}
\label{sec:DSO}

The DSO aims to ensure the reliable and cost-effective operation of the distribution grid in response to the power injections scheduled by the CEAs. Specifically, it minimizes the total procurement and dispatch costs while maintaining system-wide power balance and voltage limits. The upstream power purchase is settled at the wholesale market price, while dispatchable local generators are operated based on their marginal costs. In this model, the DSO determines the nodal DLMPs through the dual variables of the power balance constraints, which are fed back to the CEAs as price signals for subsequent scheduling. The objective function is given as:
\begin{align}
   f^{DSO} =  \sum_{t \in \mathcal{T}} \pi ^{\text{LMP}}_t\cdot p^g_{0,t} +\sum_{t \in \mathcal{T}} \sum_{l \in \mathcal{N} \setminus{\mathcal{N}_s}} c^g_{l} \cdot p^g_{l,t} 
\end{align}

To ensure a tractable convex formulation, we adopt the linearized DistFlow model \cite{baran1989network}. In this approximation, quadratic power loss terms are neglected. In addition, we impose operational limits on bus voltages and generator outputs. The resulting linearized constraints and their associated Lagrange multipliers $\lambda^\bullet$ are given below. Note that these constraints are defined for all buses \( l \in \mathcal{N} \) and time periods \( t \in \mathcal{T} \), although such indices are omitted in individual expressions for brevity.
\vspace{-1.0em}
\begin{subequations}
\label{eq:lindistflow}
\begin{align}
    &u_{k,t} = u_{l,t} - 2(r_{kl}P_{lk,t} + x_{kl}Q_{lk,t})  \qquad \qquad : \lambda^{\text{volt}}_{lk,t} \label{eq:lindistflow_voltage} \\
    &(v^\text{min})^2 \leq u_{l,t} \leq (v^\text{max})^2 \qquad  \qquad \qquad \qquad: \underline{\lambda}^u_{l,t},  \overline{\lambda}^u_{l,t}\\
    &\sum_{k:(0,k) \in \mathcal{L}} P_{0k,t} 
    - p^{\text{inj}}_{0,t} - p^g_{0,t}= 0 
    \qquad \qquad \qquad: \lambda^p_{0,t} \label{eq:lindistflow_Pbal1} 
\end{align}
\begin{align}
    &\sum_{k:(l,k) \in \mathcal{L}} P_{lk,t}-\sum_{k:(k,l) \in \mathcal{L}} P_{kl,t}  
    - p^{\text{inj}}_{l,t} -p^g_{l,t}= 0 
    : \lambda^p_{l,t} \label{eq:lindistflow_Pbal2} \\
    &\sum_{k:(0,k) \in \mathcal{L}} Q_{0k,t} 
    - q^{\text{inj}}_{0,t} - q^g_{0,t}= 0 
    \qquad \qquad \qquad: \lambda^q_{0,t} \label{eq:lindistflow_Qbal1} \\
    &\sum_{k:(l,k) \in \mathcal{L}} Q_{lk,t}-\sum_{k:(k,l) \in \mathcal{L}} Q_{kl,t}  
    - q^{\text{inj}}_{l,t} - q^g_{l,t}= 0 
    : \lambda^q_{l,t} \label{eq:lindistflow_Qbal2} \\
    &\underline{p}^g_l \leq p^g_{l,t} \leq \overline{p}^g_l   \qquad \qquad \qquad  \qquad \qquad \qquad \underline{\lambda}^{gp}_{l,t}, \ \overline{\lambda}^{gp}_{l,t} \label{eq:gen_p_limit} \\
    &\underline{q}^g_l \leq q^g_{l,t} \leq \overline{q}^g_l   \qquad \qquad \qquad  \qquad \qquad \qquad \underline{\lambda}^{gq}_{l,t}, \ \overline{\lambda}^{gq}_{l,t} \label{eq:gen_q_limit}
\end{align}
\end{subequations}

The lines' apparent power flows are neglected assuming that voltage problems dominate in the distribution system. This simplification avoids convex quadratic constraints, though the same reformulation approach applies and such an extension is straightforward. In the nodal power balance constraint~\eqref{eq:lindistflow}, power injections are related to the net active and reactive power consumption of the connected community and load forecasts as $p_{l,t}^\text{inj} =  -{p}_{l,t}^c,\forall l \in \mathcal{N}_{c}$ otherwise $p_{l,t}^\text{inj} =  -\hat{P}_l^\text{load}, \forall l \in \mathcal{N}\backslash \mathcal{N}_{c}$. Similarly, reactive power injections follow $q_{l,t}^\text{inj} =  {q}_{l,t}^c,\ \forall l \in \mathcal{N}_{c}$ and $q_{l,t}^\text{inj} =  -\hat{Q}_l^\text{load},\ \forall l \in \mathcal{N}\backslash \mathcal{N}_{c}$.

\vspace{-1.0em}
\subsection{Single-level Reformulation via KKT Conditions}
As described in Section~\ref{sec: Description_of_structure}, the interaction between the CEA and the DSO is formulated as a bilevel optimization problem, where the CEA (upper-level) determines community-level energy schedules, and the DSO (lower-level) responds with optimal dispatch and pricing decisions. To enable tractable computation, this bilevel structure is reformulated into a single-level optimization problem by replacing the lower-level DSO problem with its KKT optimality conditions \cite{boyd2004convex}. This transformation ensures that the DSO solution remains optimal while removing the nested structure of the original bilevel formulation.

Moreover, the original CEA objective \eqref{eq:cea_obj} includes bilinear terms $\lambda^p_{l,t} \cdot p^c_{l,t}$ and $\lambda^q_{l,t} \cdot q^c_{l,t}$, which increases computational complexity due to its nonlinearity. We address this by leveraging the strong duality property of the convex DSO problem to eliminate these bilinear terms and replacing them with the dual representation of the lower-level objective. The resulting objective function is given by:
\begin{align}
\label{eq:single_obj}
f^{\text{SINGLE}} = 
& \sum_{t \in \mathcal{T}} \sum_{l \in \mathcal{N}} c^{g}_{l} \cdot p^{g}_{l,t}
- \sum_{t \in \mathcal{T}} \sum_{l \in \mathcal{N} \setminus \mathcal{N}_c} \lambda^{p}_{l,t} \cdot \hat{P}^{\text{load}}_{l,t} \nonumber \\
&- \sum_{t \in \mathcal{T}} \sum_{l \in \mathcal{N} \setminus \mathcal{N}_c} \lambda^{q}_{l,t} \cdot \hat{Q}^{\text{load}}_{l,t} \nonumber + \sum_{t \in \mathcal{T}} \sum_{l \in \mathcal{N}_c} \pi^{\text{flex}} \cdot p^{\text{c,red}}_{l,t}\nonumber  \\
&+ \sum_{t \in \mathcal{T}} \sum_{l \in \mathcal{N}} \left( \overline{\lambda}^{gp}_{l,t} \cdot \overline{p}^{g}_l - \underline{\lambda}^{gp}_{l,t} \cdot \underline{p}^{g}_l 
\right) \nonumber \\
& + \sum_{t \in \mathcal{T}} \sum_{l \in \mathcal{N}} \left( \overline{\lambda}^{gq}_{l,t} \cdot \overline{q}^{g}_l - \underline{\lambda}^{gq}_{l,t} \cdot \underline{q}^{g}_l 
\right)  \nonumber  \\
&+ \sum_{t \in \mathcal{T}} \sum_{l \in \mathcal{N}} \left( 
\overline{\lambda}^{u}_{l,t} \cdot (v^{\max})^2 
- \underline{\lambda}^{u}_{l,t} \cdot (v^{\min})^2 
\right)
\end{align}

Note that all terms in the objective function are linear as a result of the dual substitution. 
Together with the linear or linearized constraints described below, this yields a tractable optimization model.

The single-level reformulation is subject to all primal constraints of the CEA and DSO models, including community-level DER scheduling constraints~\eqref{eq:cea_constraints} and distribution network power flow equations~\eqref{eq:lindistflow}. 
To ensure the optimality of the lower-level DSO problem, the KKT conditions — in addition to the primal feasibility constraints — are imposed as additional constraints in the single-level reformulation. These conditions consist of three components.

First, the \textit{stationarity conditions} require that the gradient of the DSO Lagrangian \(\mathcal{L}(x, \lambda)\), with respect to the DSO’s primal variables \(x\), must be zero:
\vspace{-0.3em}
\begin{subequations}
\label{eq:kkt_conditions}
\begin{align}
&\nabla_{x} \mathcal{L}(x, \lambda) = 0 \label{eq:kkt_stationarity}
\end{align}

Second, the \textit{dual feasibility conditions} require all dual variables \(\lambda^\bullet\) for inequality constraints to be non-negative:
\vspace{-0.3em}
\begin{align}
&\lambda^\bullet \geq 0 \label{eq:kkt_dual_feasibility}
\end{align}

Third, the \textit{complementarity slackness conditions} are linearized using a standard Big-M reformulation \cite{fortuny1981representation}. For each inequality constraint \(c_i(x) \leq 0\) and its associated dual variable \(\lambda_i\), a binary variable \(z_i \in \{0,1\}\) is introduced:
\vspace{-0.1em}
\begin{align}
&\lambda_i \leq \mathfrak{M} z_i, \quad c_i(x) \leq \mathfrak{M}(1 - z_i) \label{eq:kkt_complementarity}
\end{align}
\end{subequations}

The final single-level optimization problem, structured as a Mixed Integer Linear Program (MILP), is:
\begin{equation}
\label{eq:single_level}
\begin{aligned}
\min \quad & f^{\text{SINGLE}}~\eqref{eq:single_obj} \\
\text{subject to} \quad 
& \text{CEA constraints: } \eqref{eq:cea_constraints}, \\
& \text{DSO KKT conditions: } \eqref{eq:lindistflow}, \eqref{eq:kkt_conditions}
\end{aligned}
\end{equation}

The overall reformulation process eliminates the nested bilevel structure by imposing the KKT conditions for the DSO problem, while the strong duality theorem is used to replace the bilinear objective of the CEA with the dual of the DSO. This reformulation procedure is visually summarized in Fig.~\ref{fig:cea_dso_framework}.
\vspace{-1.5em}

\section{Coalitional Game and\\ Shapley Value Allocation}
\label{sec:allocation}

The optimal value obtained by solving \eqref{eq:single_level} represents the total operational cost aggregated over all participating ECs. Although this joint scheduling ensures overall cost minimization from the CEA's perspective, it does not guarantee fairness in allocating costs to individual ECs. Under this centralized coordination scheme, the CEA is granted full control over DERs owned by the ECs, and optimizes their collective operation to minimize the total cost across all ECs. As a result, the resources of a specific EC may be dispatched in ways that primarily benefit others, potentially leading to cost allocations that do not accurately reflect each EC’s contribution. If costs are allocated solely based on nodal energy usage under the jointly optimized schedule, the resulting allocation may not fully capture each EC’s contribution, and could result in unfair cost allocations.

To address this issue, we model the cost allocation problem as a coalitional game in characteristic form and adopt the Shapley value as a solution concept to improve fairness in allocating the collective cost savings based on each EC’s marginal contribution.
Let \(\mathcal{M}\) denote the set of participating ECs, and let \(\upsilon : 2^{\mathcal{M}} \to \mathbb{R}\) be the characteristic function that maps each coalition to the corresponding cost savings achieved through cooperation\cite{osborne1994course}. The individual cost \(c_m^{\text{indiv}}\) of EC \(m\) is computed by solving problem~\eqref{eq:single_level} under a modified setting in which only EC \(m\) participates, while all other ECs’ DER schedules (generation, load, storage) are fixed at their base-case profiles. Similarly, the cooperative cost \(c_{\mathcal{C}}\) for a coalition \(\mathcal{C} \subseteq \mathcal{M}\) is obtained by solving the same optimization problem for the participating ECs in \(\mathcal{C}\), while treating non-participating ECs as passive entities 
\cite{chics2017coalitional}.

The value of a coalition \(\mathcal{C}\) can then be defined as the total cost savings achieved by the coalition relative to operating individually:
\begin{equation}
    \upsilon(\mathcal{C}) = \sum_{m \in \mathcal{C}} c_m^{\text{indiv}} - c_{\mathcal{C}}
    \label{eq:value_function}
\end{equation}
This definition ensures that \(\upsilon(\emptyset) = 0\) and \(\upsilon(\mathcal{M})\) corresponds to the total cost savings of the grand coalition.

In cooperative games with transferable utility, the value of a coalition must be distributed among its members according to a fair allocation rule. Among various solution concepts proposed in cooperative game theory, the Shapley value provides a unique and axiomatic method based on each player’s average marginal contribution \cite{shapley1953value}. In our context, it is used to allocate the total cost savings \(\upsilon(\mathcal{M})\) of the grand coalition to individual communities in a manner that reflects their contribution to cooperative performance.

For a set of communities \(\mathcal{M}\) with \(|\mathcal{M}| = N\), the Shapley value assigned to each community \(m \in \mathcal{M}\) is given by:
\begin{equation}
    \phi_m(\upsilon) = \sum_{S \subseteq \mathcal{M} \setminus \{m\}} \frac{|S|!(N - |S| - 1)!}{N!} \left[ \upsilon(S \cup \{m\}) - \upsilon(S) \right]
    \label{eq:shapley}
\end{equation}
Here, the term \(\upsilon(S \cup \{m\}) - \upsilon(S)\) represents the marginal contribution of community \(m\) to subset \(S\). The sum of all Shapley values satisfies the efficiency property, i.e., \(\sum_{m \in \mathcal{M}} \phi_m(\upsilon) = \upsilon(\mathcal{M})\). In our framework, the total cost savings from joint optimization is therefore fully and fairly allocated to participating communities based on their Shapley values.

Intuitively, the Shapley value assigns a larger share of the total cost savings to communities that consistently make greater marginal contributions across different coalitions. For instance, if the BESS of a specific community plays a key role in reducing the total operational cost of the cooperating communities, that community will receive a higher Shapley value. This allocation reflects the synergistic benefit of integrating its resources into the cooperative operation.

Note that the independence of optimization problems for different coalitions allows the Shapley value to be computed in parallel, improving scalability. However, the number of coalitions grows exponentially with the number of ECs, which makes exact computation intractable for large systems. We address this with a signature-based approximation method that efficiently samples a representative subset of coalitions. Details and numerical validation are in Section IV.

After allocating the cooperative cost savings via the Shapley value, the final cost incurred by each community is obtained by subtracting its share of the cost savings from its individual cost:
\begin{equation}
    c_m^{\text{final}} = c_m^{\text{indiv}} - \phi_m(\upsilon)
    \label{eq:final_cost}
\end{equation}
\section{Numerical Validation}
\label{sec:numerical_validation}
\subsection{Simulation Setup}
\begin{figure}[!htbp]
    \centering
        \includegraphics[width=1.0\linewidth]{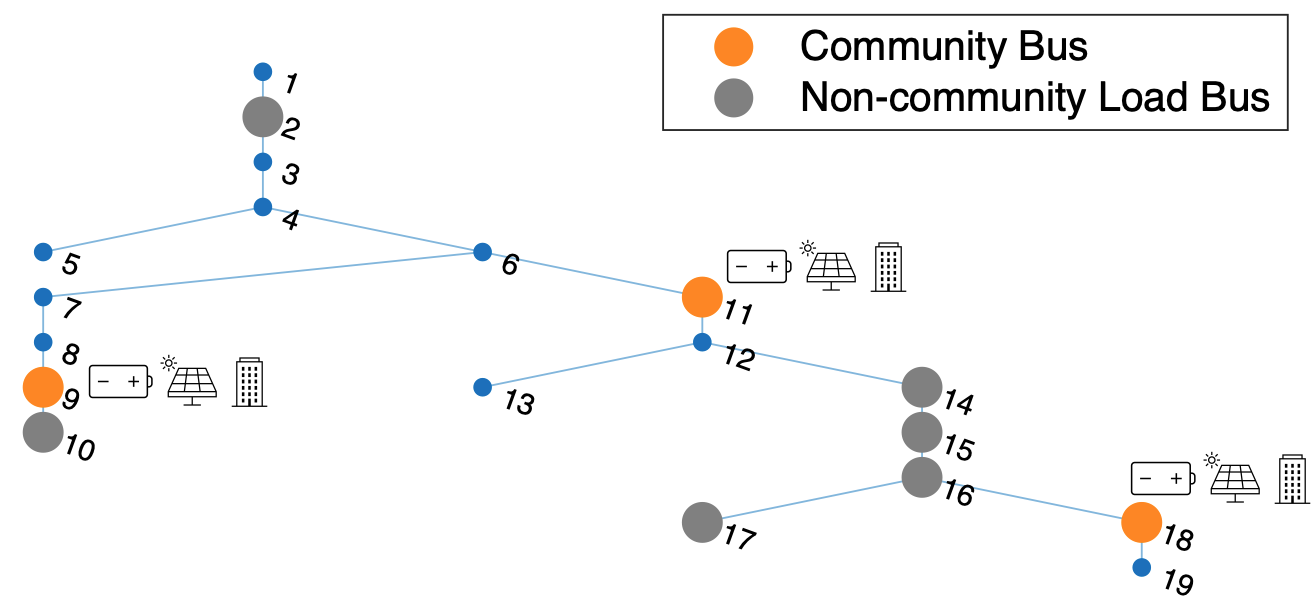}
    \vspace{-2.5em}
    \caption{CIGRE low voltage benchmark network.}
    \label{fig:cigre}
\end{figure}

To evaluate the effectiveness of the proposed fair allocation mechanism based on the Shapley value, numerical simulations are performed on various distribution networks under realistic operating conditions. Fig. \ref{fig:cigre} illustrates the benchmark CIGRE low-voltage distribution network used in this study\cite{CIGREREF}. The feeder is connected to the upstream transmission system via a 20/0.4 kV, 400 kVA transformer. The system includes both community and non-community loads, and the impact of community locations on system-level fairness and grid operation is analyzed under multiple scenarios.

Each energy community is composed of DERs, including PV systems, BESS, and controllable loads. It is assumed that the communities are allowed to adjust up to 30\% of their base demand through flexible load response, with a compensation rate of \$75/MWh \cite{upadhaya2019optimal}. The electricity price at the transmission-distribution interface is based on the 2024 average of hourly CAISO locational marginal prices (LMP) for the NP-15 zone, which ranges approximately from \$12/MWh to \$55/MWh
 \cite{caiso_oasis}. In addition, each distribution node is assumed to have a local dispatchable generator operating at a marginal cost of \$250/MWh, based on values reported in \cite{mateo2020comprehensive}. The reactive power capabilities of dispatchable generators and PV inverters are constrained by typical grid code requirements. Generator output is limited by a minimum power factor of 0.9, which corresponds to a reactive power limit of approximately 0.484 times the active power, while PV inverters are restricted to supplying reactive power up to 20\% of their real power output \cite{jangid2024distribution}. The charging and discharging efficiencies of BESS are both set to 95\%, i.e., $\eta^{\text{ch}}_l = \eta^{\text{dis}}_l = 0.95$, and simultaneous charging and discharging are not permitted. The initial and final state of charge (SoC) is set to 50\% of the rated capacity.

We model the load and PV generation using the real measurements from the smart grid setup at the EPFL, Switzerland \cite{gupta2020grid}. The data\footnote{\url{https://github.com/DESL-EPFL/DESL-Photovoltaic-timeseries}}, which is available at per second resolution, is down-sampled to hourly resolution for our simulation.

To test scalability under different network sizes, we also conducted additional simulations on larger distribution systems, specifically
case69\cite{baran1989network} and case123\cite{cikan2024reconfiguration}. Across all test cases, the same modeling assumptions and DER parameters are applied. Simulations are implemented in MATLAB using YALMIP with Gurobi as the solver~\cite{gurobi2025}, and executed on a laptop with 24GB memory.
\vspace{-0.5em}

\subsection{Simulation Results}
\subsubsection{CIGRE 19-bus system}
\label{sec:19_bus}

We consider a scenario based on the CIGRE 19-bus test system, where three communities are located at Bus 9, Bus 11, and Bus 18. Each community is initially assumed to have an identical DER setup consisting of PV, controllable demand, and a BESS rated at 20~kW/50~kWh.

For each community, three cost-related quantities are evaluated. The individual cost $c_m^{\text{indiv}}$ is obtained by solving problem~\eqref{eq:single_level} with only that community active while fixing the net injections of the others. This represents the cost if each community operates independently without coordination. The Shapley-based saving $\phi_m(\upsilon)$ is computed using \eqref{eq:shapley} from the costs of all possible coalitions and quantifies the marginal contribution of community~$m$ to the grand coalition. Finally, the final settlement for each community is calculated as $c_m^{\text{final}} = c_m^{\text{indiv}} - \phi_m(\upsilon)$ according to \eqref{eq:final_cost}, ensuring that the cost reduction from cooperation is allocated based on the contribution of each participant. 

In addition to the Shapley-based allocation (referred to as the \textsf{Shapley} method), we also consider a baseline method denoted by \textsf{Base}. In this approach, each community's cost, $c_m^{\text{base}}$, is calculated directly from the DLMPs and dispatch results from the solution to problem \eqref{eq:single_level}, without accounting for the community's marginal contribution to cost savings. The calculation is given by:
\begin{equation*}
c_m^{\text{base}} = \sum_{t \in \mathcal{T}} \left( \lambda^p_{m,t} \cdot p^c_{m,t} + \lambda^q_{m,t} \cdot q^c_{m,t} + \pi^{\text{flex}} \cdot p^{\text{c,red}}_{m,t} \right).
\end{equation*}

\begin{figure}[b]
    \centering
    \includegraphics[width=1.0\linewidth]{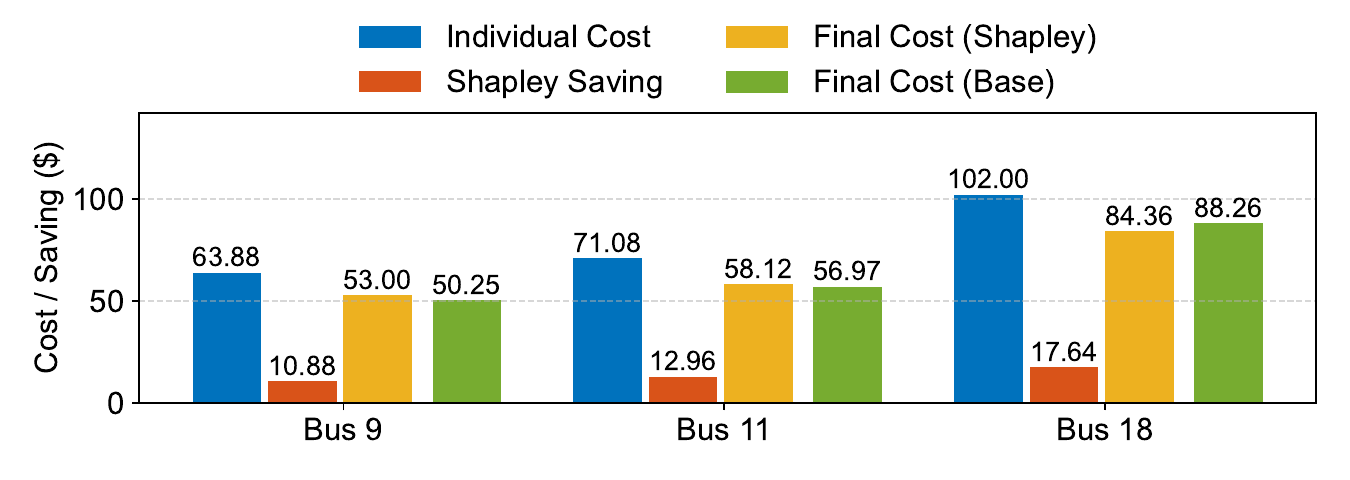}
    \vspace{-2.5em}
    \caption{Individual cost, Shapley-based savings, and final cost for each community bus in the CIGRE system, under Shapley and base allocation methods. The Shapley-based final cost reflects each community's marginal contribution, while the base method directly applies DLMPs and dispatch results without fairness considerations.}
    \label{fig:cost_plot_19bus}
\end{figure}

Fig.~\ref{fig:cost_plot_19bus} compares the individual costs, the Shapley-based savings, and the final allocated costs 
under the \textsf{Shapley} and \textsf{Base} methods. Even with identical DER capacities, the resulting costs differ across communities due to their locations and their impact on overall system operation. All communities achieve a lower cost than in the individual case, showing that cooperation through the CEA reduces the total cost. 
While both methods produce the same total system cost, the \textsf{Shapley} method distributes the savings based on marginal contributions, whereas the \textsf{Base} method directly uses DLMP results without considering individual contributions.

\begin{figure}[!htbp]
    \centering
    \includegraphics[width=1.0\linewidth]{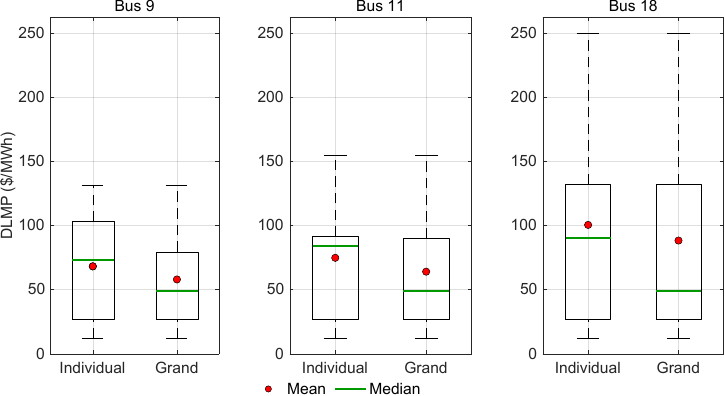}
    \vspace{-1.5em}
    \caption{Active power DLMP distributions at the community nodes under individual activation and grand coalition participation.}
    \vspace{-1.5em}
    \label{fig:DLMP_plot}
\end{figure}

Among the three, the community located at Bus 18 exhibits the highest individual and final costs, along with the largest Shapley value. This result can be further explained by Fig.~\ref{fig:DLMP_plot}, which shows the active power DLMP distributions at each community node under individual and grand coalition scenarios. Due to voltage constraints, the DLMPs at Bus 18 are significantly higher, making its DERs more influential in alleviating local constraints. As a result, its contribution to reducing overall system costs is the most significant, which is reflected in its higher Shapley value. Across all buses, DLMP distributions shift lower under the grand coalition, indicating that coordinated participation reduces the marginal cost of supplying power and benefits all communities.

\begin{table}[t]
\centering
\caption{Final cost comparison under BESS expansion (3× original capacity) at a single community. ``Comm.'' denotes the bus ID of each community.}
\label{tab:bess3x_centered}
\renewcommand{\arraystretch}{1.2}
\begin{tabular}{lccccc}
\hline
\textbf{Method} & \textbf{Scenario} & \textbf{Comm.} & \begin{tabular}[c]{@{}c@{}}Original \\ Cost (\$)\end{tabular} & 
\begin{tabular}[c]{@{}c@{}} BESS x3 \\ Cost (\$)\end{tabular} & 
\begin{tabular}[c]{@{}c@{}}Cost \\ Diff. (\$)\end{tabular} \\
\hline
\multirow{12}{*}{\textsf{Shapley}} 
& BESS@9 x3  & \textbf{Bus 9}  & \textbf{53.00} & \textbf{39.31} & \textbf{13.69} \\
&            & Bus 11         & 58.12 & 58.21 & $-$0.08 \\
&            & Bus 18         & 84.36 & 82.62 & 1.75 \\
& \textbf{Total} &             & \textbf{195.48} & \textbf{180.13} & \textbf{15.35} \\
\cline{2-6}
& BESS@11 x3 & Bus 9          & 53.00 & 49.15 & 3.85 \\
&            & \textbf{Bus 11} & \textbf{58.12} & \textbf{40.66} & \textbf{17.46} \\
&            & Bus 18         & 84.36 & 80.61 & 3.75 \\
& \textbf{Total} &             & \textbf{195.48} & \textbf{170.42} & \textbf{25.06} \\
\cline{2-6}
& BESS@18 x3 & Bus 9          & 53.00 & 47.77 & 5.23 \\
&            & Bus 11         & 58.12 & 52.17 & 5.95 \\
&            & \textbf{Bus 18} & \textbf{84.36} & \textbf{54.86} & \textbf{29.50} \\
& \textbf{Total} &             & \textbf{195.48} & \textbf{154.80} & \textbf{40.68} \\
\hline
\multirow{12}{*}{\textsf{Base}} 
& BESS@9 x3  & \textbf{Bus 9}  & \textbf{50.25} & \textbf{38.82} & \textbf{11.43} \\
&            & Bus 11         & 56.97 & 56.62 & 0.35 \\
&            & Bus 18         & 88.26 & 84.69 & 3.57 \\
& \textbf{Total} &             & \textbf{195.48} & \textbf{180.13} & \textbf{15.35} \\
\cline{2-6}
& BESS@11 x3 & Bus 9          & 50.25 & 46.65 & 3.61 \\
&            & \textbf{Bus 11} & \textbf{56.97} & \textbf{45.93} & \textbf{11.04} \\
&            & Bus 18         & 88.26 & 77.84 & 10.42 \\
& \textbf{Total} &             & \textbf{195.48} & \textbf{170.42} & \textbf{25.06} \\
\cline{2-6}
& BESS@18 x3 & Bus 9          & 50.25 & 41.89 & 8.36 \\
&            & Bus 11         & 56.97 & 46.63 & 10.34 \\
&            & \textbf{Bus 18} & \textbf{88.26} & \textbf{66.28} & \textbf{21.98} \\
& \textbf{Total} &             & \textbf{195.48} & \textbf{154.80} & \textbf{40.68} \\
\hline
\end{tabular}
\vspace{-0.5em}
\end{table}

\begin{table}[t]
\centering
\caption{Final cost comparison when expanding BESS at Bus 11 to 3.5$\times$ its original capacity. ``Comm.'' denotes the bus ID of each community.}
\label{tab:bess35x}
\renewcommand{\arraystretch}{1.2}
\begin{tabular}{lccccc}
\hline
\textbf{Method} & \textbf{Scenario} & \textbf{Comm.} & \begin{tabular}[c]{@{}c@{}}Original \\ Cost (\$)\end{tabular} & 
\begin{tabular}[c]{@{}c@{}}BESS x3.5 \\ Cost (\$)\end{tabular} & 
\begin{tabular}[c]{@{}c@{}}Cost \\ Diff. (\$)\end{tabular} \\
\hline
\multirow{4}{*}{\textsf{Shapley}} 
& BESS@11  & Bus 9   & 53.00 & 47.89 & 5.11 \\
&  x3.5              & \textbf{Bus 11}  & \textbf{58.12} & \textbf{36.41} & \textbf{21.71} \\
&               & Bus 18  & 84.36 & 79.43 & 4.93 \\
& \textbf{Total} &         & \textbf{195.48} & \textbf{163.73} & \textbf{31.76} \\
\hline
\multirow{4}{*}{\textsf{Base}} 
& BESS@11   & Bus 9   & 50.25 & 44.44 & 5.81 \\
& x3.5              & \textbf{Bus 11}  & \textbf{56.97} & \textbf{44.19} & \textbf{12.78} \\
&               & Bus 18  & 88.26 & 75.09 & 13.17 \\
& \textbf{Total} &         & \textbf{195.48} & \textbf{163.73} & \textbf{31.76} \\
\hline
\end{tabular}
\vspace{-1.5em}
\end{table}

\textit{Sensitivity with BESS size:} To examine fairness concerns that may arise when allocating the outcome of problem~\eqref{eq:single_level} directly to each community (i.e., when using the \textsf{Base} method), we consider scenarios in which the BESS capacity at a single site is selectively expanded. 
Table~\ref{tab:bess3x_centered} presents the final cost of each community when the BESS at one site is increased to three times its original capacity. In all cases, the additional BESS enhances operational flexibility and enables peak shaving, leading to reduced system-wide costs. However, how these savings are distributed across communities differs significantly depending on the allocation rule.

Notably, the \textsf{Shapley} allocation more precisely assigns the resulting cost savings to the community that contributes additional BESS capacity. For example, when Bus 11 hosts the expanded BESS, the cost at Bus 11 decreases from \$58.12 to \$40.66 (a reduction of \$17.46) under \textsf{Shapley} allocation. In contrast, under the \textsf{Base} allocation, the cost at Bus 11 decreases from \$56.97 to \$45.93 (\$11.04 reduction), while Bus 18 also benefits with a cost decrease from \$88.26 to \$77.84 (\$10.42 reduction), despite not contributing any additional resources. This highlights a key limitation of the baseline method in capturing localized benefits.

Table \ref{tab:bess35x} further extends the analysis by considering a case where the BESS capacity at Bus 11 is expanded to 3.5 times its original size. Under the \textsf{Shapley} method, the total cost decreases from \$195.48 to \$163.73, with the largest benefit concentrated at Bus 11, whose cost drops from \$58.12 to \$36.41 (\$21.71 reduction).
In contrast, under the \textsf{Base} method, the cost at Bus 11 decreases from \$56.97 to \$44.19 (\$12.78 reduction), while Bus 18—without any new investment—sees its cost drop from \$88.26 to \$75.09 (\$13.17 reduction), which is even larger. This result highlights a potential limitation of the \textsf{Base} allocation—it may not fully reflect the origin of the cost savings, which could result in a less fair allocation. Such misalignment between contribution and benefit may weaken the incentive for communities to invest in grid-supportive resources like BESS.

Even in the one case where a community's allocated cost increases compared to the original allocation—namely, Bus~11 under the ``BESS@9 x3'' scenario—the \textsf{Shapley} allocation remains cooperative. Although Bus~11's cost rises slightly to \$58.21, it is still well below its individual cost of \$71.08 (Fig.~\ref{fig:cost_plot_19bus}), ensuring no incentive to leave the coalition. This outcome upholds the principle of individual rationality, as all communities benefit from cooperation relative to acting alone.

\begin{figure}[!htbp]
    \centering
    \includegraphics[trim=0.0cm 0.05cm 0.05cm 0.0cm, clip,width=0.9\linewidth]{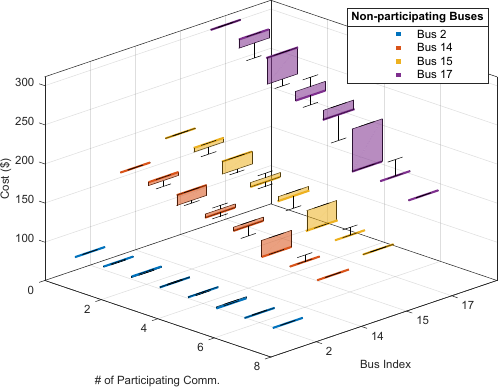}    
\caption{3D box-plot of non-participating buses’ cost distribution across different numbers of participating communities.}
    \label{fig:cost_regulator}
\end{figure}

\textit{Sensitivity with the number of participating communities:} To examine the impact on buses that are not part of the coalition, we analyzed how the share of the system managed by the CEA affects non-community nodes when more communities participate in coordinated operation. Specifically, in the same distribution system, communities are located at buses 4, 6, 7, 9, 11, 12, and 18, and the costs of non-participating nodes 2, 10, 14, 15, 16, and 17 are evaluated. Fig. \ref{fig:cost_regulator} shows the cost distribution for non-participating buses with number of participating communities, highlighting both the general downward trend and the variability caused by different coalition compositions. The results show that the average cost of non-community load buses decreases as more communities join the coalition and the CEA optimizes a larger share of the system, indicating that lowering DLMPs at the system level also benefits other buses. However, even with the same number of participating communities, noticeable variation in costs is observed depending on the specific combination of communities in the coalition. This indicates that increasing the number of participating communities does not necessarily guarantee benefits for all non-participating buses, as the specific coalition composition can lead to widely varying outcomes. For clarity, results for buses 10 and 16 are omitted, but they exhibit similar trends.

\vspace{-1.9em}
\begin{figure}[!htbp]
    \centering
    \includegraphics[trim=2.3cm 0.1cm 0.1cm 0.1cm, clip, width=0.8\linewidth]{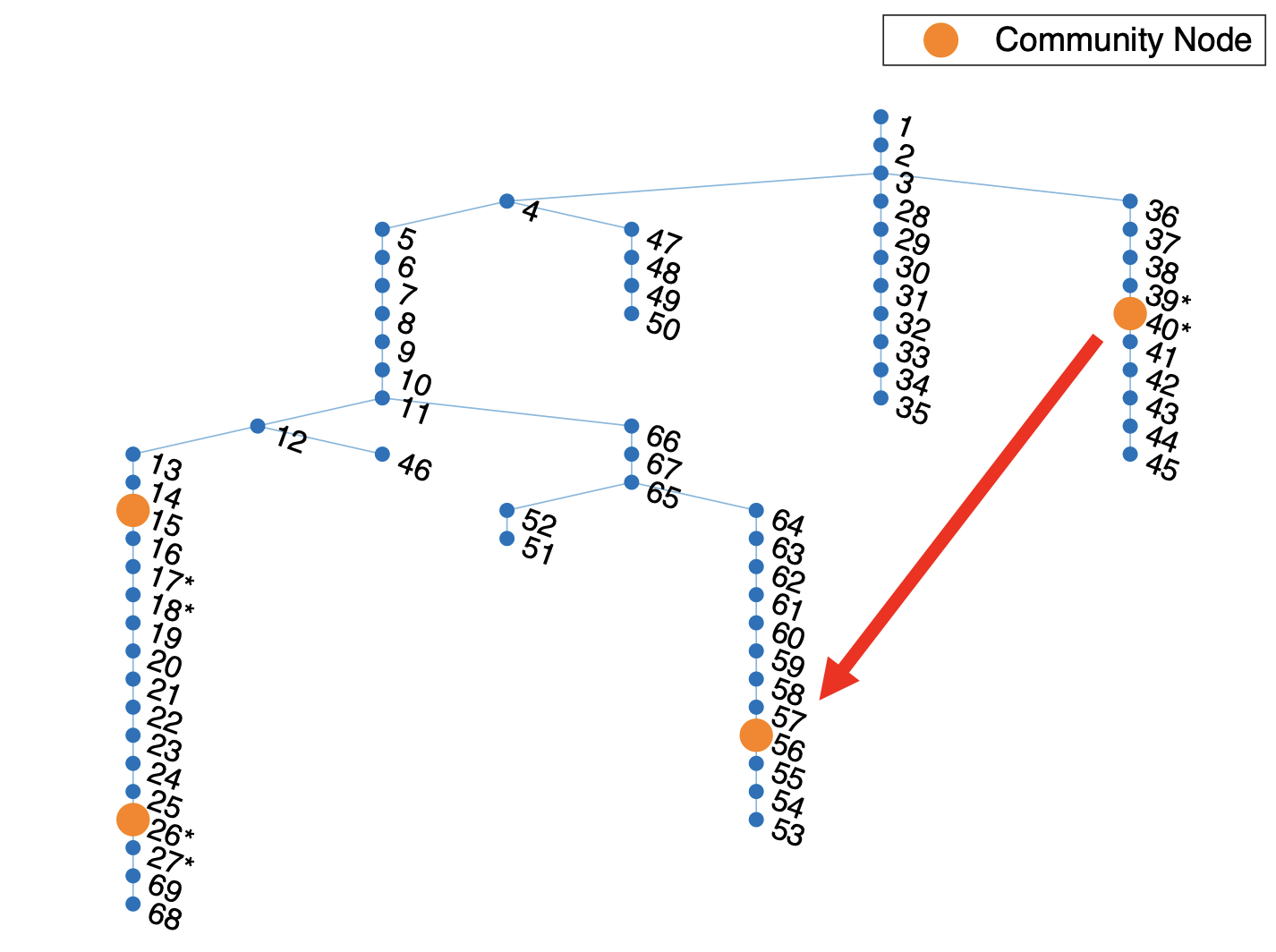}
    \vspace{-1.0em}
    \caption{IEEE 69-bus distribution system showing three different EC placement configurations.
    The first two configurations consider three ECs located at buses 15, 26, and 40 
    and at buses 15, 26, and 56, respectively. 
    The third configuration places six ECs at Buses 17, 18, 26, 27, 39, and 40, marked with an asterisk (\textasteriskcentered), to demonstrate the signature-based approximation. Non-community load buses are omitted for clarity.}
    \label{fig:bus69_star}
\end{figure}

\begin{figure}[!htbp]
    \centering
    \includegraphics[width=1.0\linewidth]{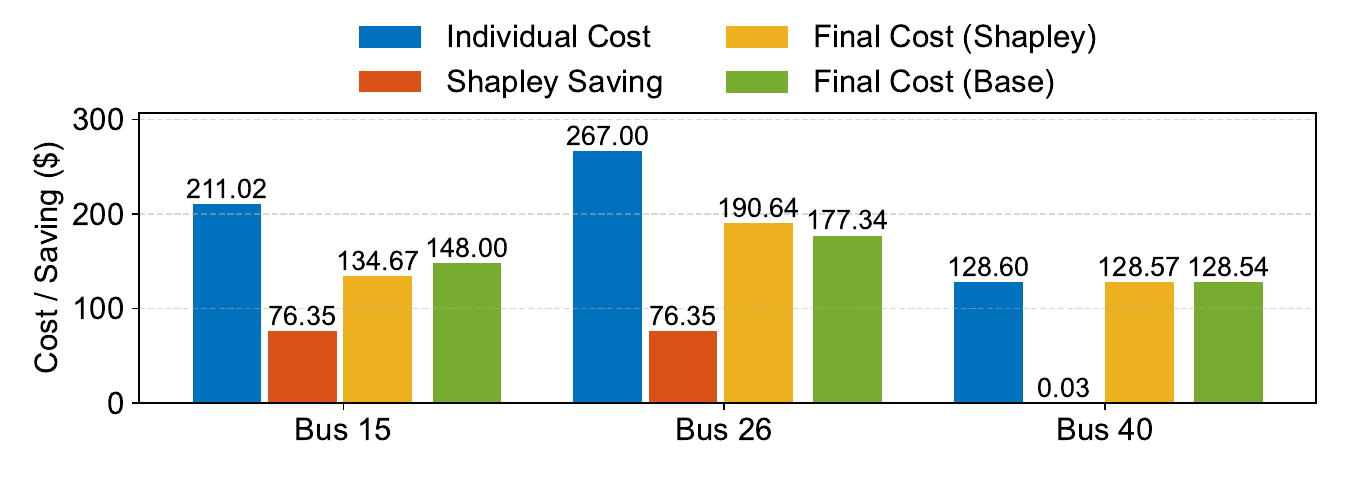}
    \vspace{-2.5em}
\caption{Individual cost, Shapley-based savings, and final cost for each community node in the IEEE 69-bus system (ECs at Buses 15, 26, 40), under \textsf{Shapley} and \textsf{Base} allocation methods.}
    \label{fig:cost_plot_69bus}
    \vspace{-1.5em}
\end{figure}
\subsubsection{IEEE 69-bus system}
\label{sec:69_bus}
We extend the analysis to the IEEE 69-bus test system, and we consider multiple scenarios depicting different placement of ECs to examine the impact of community location on cost savings. All ECs are assumed to have identical DER portfolios and capacities. Again, we compare \textsf{Shapley} and \textsf{Base} methods as defined earlier.

We first examine a configuration where one community is placed farther away from the others, with communities located at Buses 15, 26, and 40. The results are visualized in Fig.~\ref{fig:cost_plot_69bus}. As observed, the community at Bus 40 has a similar costs for individual and cooperative case, illustrating minimal influence on overall CEA operation. This may be due to weak coupling with the rest of the network. In contrast, the communities at Buses 15 and 26 exhibit equal Shapley savings, reflecting their symmetric contributions to system-wide cost savings. This symmetry is not preserved under the \textsf{Base} method, which leads to a relatively more favorable allocation for Bus 26, illustrating the base method’s inability to fully reflect contribution-based fairness. Nevertheless, the contributing communities still benefit from participating in the coalition, with final costs lower than their individual operation costs, preserving the principle of individual rationality. 

\begin{figure}[!b]
    \centering
    \includegraphics[width=1.0\linewidth]{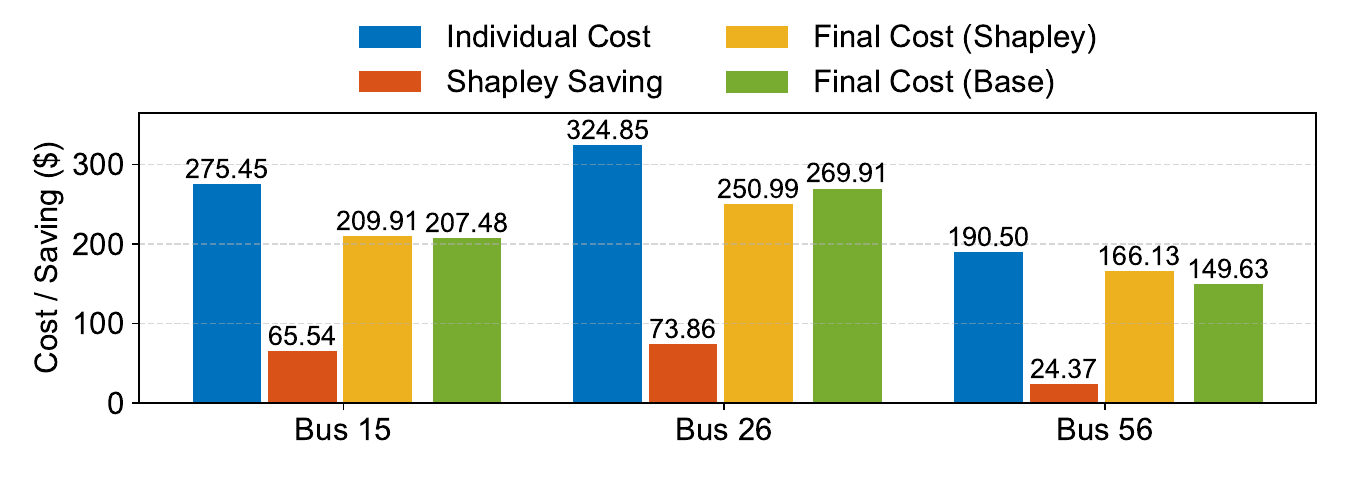}
    \vspace{-2.5em}
\caption{Individual cost, Shapley-based savings, and final cost for each community node in the IEEE 69-bus system (ECs at Buses 15, 26, 56), under \textsf{Shapley} and \textsf{Base} allocation methods.}

    \label{fig:cost_plot_69bus_2}
\end{figure}
\begin{figure}[!b]
    \centering
    \includegraphics[width=1.0\linewidth]{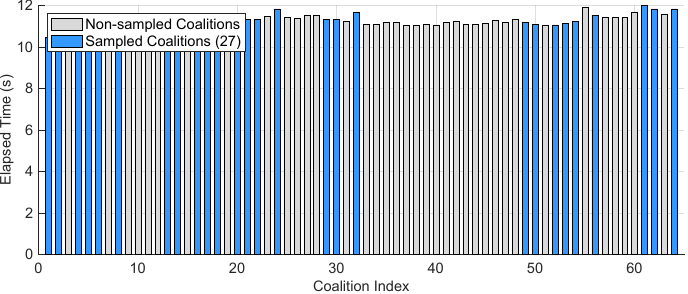}
    \vspace{-2.0em}
\caption{Elapsed time for computing all 64 coalitions. Sampled coalitions (blue) correspond to the 27 representative signatures.}
    \label{fig:comp_time_bar}
\end{figure}

Next, we consider a configuration in which the community at Bus 40 is relocated to Bus 56 to consider the impact of location on the cost saving. The computed Shapley values are accordingly non-zero for all three communities, as can be seen in Fig.~\ref{fig:cost_plot_69bus_2}. In this configuration, all three communities interact more strongly via grid constraints, and each contributes non-negligibly to the coalition’s performance. These findings highlight that the \textsf{Shapley} allocation accurately captures the locational differences in impact and ensures fair distribution across diverse network configurations.

To investigate computational scalability, we expand the study to six ECs by placing three closely located pairs at buses 17, 18, 26, 27, 39, and 40, as shown in Fig. \ref{fig:bus69_star}. It is assumed that all communities have identical DER configurations and capacities. Six ECs results in $2^6 = 64$ coalitions. Since adding more ECs results in combinatorial explosion of coalitions, we therefore propose a ``signature-based approximation''.

Instead of evaluating every possible coalition, we group together those that have a similar impact on the coalition—such as when two nearby communities can be interchanged without changing the overall result—and represent each group with a single “signature.” Formally, we define this community similarity based on two key criteria grounded in network physics: (1) electrical proximity within the network topology and (2) the portfolio of DER assets. This approach can greatly reduce the number of optimization problems that must be solved while preserving the accuracy of the resulting Shapley values.

For example, a coalition including Bus 17 produces nearly the same impact as one including Bus 18, because the two buses are located close together and contribute symmetrically to the system, so only one representative coalition needs to be evaluated. In other words, a coalition formed by communities at Buses 17, 26, and 39 yields almost the same result as the one with Buses 18, 26, and 39, so it is sufficient to evaluate only one of them. This yields $3^3 = 27$ unique group-level signatures, significantly reducing the number of coalitions to evaluate compared to the original $2^6 = 64$. 

Each of the 27 representative coalitions is solved once, and the resulting values are assigned to coalitions with same signature. Fig.~\ref{fig:comp_time_bar} compared the elapsed time for solving each coalition, highlighting the sampled subset. Table~\ref{tab:shapley_compare} compares the Shapely savings computed with exact and approximated method. Despite using fewer samples, the approximated Shapley values closely match the exact values, validating its effectiveness without significantly sacrificing fairness accuracy.

\begin{table}[t]
\centering
\caption{Comparison of exact and approximated Shapley saving\\ (IEEE 69-bus).}
\vspace{-0.5em}
\label{tab:shapley_compare}
\renewcommand{\arraystretch}{1.2}
\begin{tabular}{cccccc}
\hline
\textbf{Bus} & \textbf{Exact (\$)} & \textbf{Approx. (\$)} & \textbf{Abs. Error (\$)} & \textbf{Rel. Error (\%)} \\
\hline
17 & 31.261 & 31.321 & 0.060 & 0.19 \\
18 & 31.271 & 31.321 & 0.050 & 0.16 \\
26 & 34.096 & 34.171 & 0.075 & 0.22 \\
27 & 34.356 & 34.171 & 0.185 & 0.54 \\
39 & 0.00153 & 0.00151 & 0.00002 & 1.27 \\
40 & 0.00144 & 0.00151 & 0.00007 & 5.01 \\
\hline
\textbf{Total} & 130.987 & 130.987 & -- & -- \\
\hline
\end{tabular}
\vspace{-1.5em}
\end{table}

\begin{table}[t]
\centering
\caption{Computation time comparison (IEEE 69-bus).}
\vspace{-0.5em}
\label{tab:time_summary}
\renewcommand{\arraystretch}{1.2}
\begin{tabular}{lcc}
\hline
\textbf{Method} & \textbf{Coalitions Solved} & \textbf{Total Time (s)} \\
\hline
Full enumeration & 64 & 719 \\
Signature-based sampling & 27 & 304 \\
\hline
\end{tabular}
\vspace{-2.5em}
\end{table}

\subsubsection{IEEE 123-bus system}
\label{sec:123_bus}
To evaluate the scalability and robustness of the proposed cost allocation method, we apply the framework to the IEEE 123-bus test system. This system includes six communities with different DER capacities located at Buses 4, 6, 55, 59, 90, and 94. Unlike the 69-bus case, these communities are not directly adjacent, but they are located along the same major branch of the distribution feeder. The six communities include both identical and non-identical DER capacities to reflect diverse sizing scenarios. It is assumed that Buses 4 and 6 are identical in size, forming a symmetric pair. Bus 55 has a 20\% larger capacity than Bus 59, while Bus 94 has 50\% more DER capacity than Bus 90.

Even when DER capacities differ across communities, the final cost under \textsf{Shapley} allocation can be approximated under the assumption that it scales linearly with DER size, since communities with similar setups tend to have costs (or benefits) that are roughly proportional to their installed DER capacities. This enables us to reduce the number of coalitions by grouping similar communities. For instance, we treat the asymmetric pair (e.g., Buses 55 and 59) as symmetric during coalition sampling. 
For each community, we first compute the individual cost and the associated Shapley savings, and then obtain the final cost as their difference. 
The final costs are subsequently scaled using a post-processing adjustment ratio (e.g., 1.2:1 for Bus 55 and 59 as Bus 55 has 20\% larger capacity) to reflect the actual DER size differences, yielding reasonably accurate allocations as shown in Table~\ref{tab:shapley_compare_123}. As the results demonstrate, the approximation produces shapely savings with reasonable accuracies.

The proposed strategy is particularly feasible in practice because cost allocation is performed ex-post. That is, exact values are not required during real-time operation, and final settlements can be computed using stored data. Also, note that the total settlement amount across all communities remains unchanged, and only the allocation among participants needs to be fine-tuned afterward. That is, although individual cost allocations differ slightly between the exact and approximate methods, the total settlement is preserved, thereby maintaining efficiency. In addition, the method can be extended to further reduce computational complexity when more than two similar communities are present. For instance, even with six communities, if they can be organized into two symmetric triplets, the number of required coalitions can be reduced from 27 to 16, highlighting the method’s potential for further scaling.

\begin{table}[t]
\centering
\caption{Comparison of exact and approximated final costs (Shapley) (IEEE 123-bus).}
\vspace{-0.5em}
\label{tab:shapley_compare_123}
\renewcommand{\arraystretch}{1.2}
\begin{tabular}{ccccc}
\hline
\textbf{Bus} & \textbf{Exact (\$)} & \textbf{Approx. (\$)} & \textbf{Abs. Error (\$)} & \textbf{Rel. Error (\%)} \\
\hline
4   & 12.12  & 11.72  & 0.40  & 3.30 \\
6   & 12.12  & 11.72  & 0.40  & 3.30 \\
55  & 41.34  & 41.99  & 0.65  & 1.57 \\
59  & 50.17  & 50.39  & 0.22  & 0.44 \\
90  & 24.97  & 24.46  & 0.51  & 2.04 \\
94  & 36.24  & 36.69  & 0.45  & 1.24 \\
\hline
\textbf{Total} & 176.95 & 176.95 & -- & -- \\
\hline
\end{tabular}
\vspace{-0.5em}
\end{table}
\vspace{-0.5em}
\begin{table}[t]
\centering
\caption{Computation time comparison (IEEE 123-bus).}
\vspace{-0.5em}
\label{tab:time_summary_123}
\renewcommand{\arraystretch}{1.2}
\begin{tabular}{lcc}
\hline
\textbf{Method} & \textbf{Coalitions Solved} & \textbf{Total Time (s)} \\
\hline
Full enumeration & 64 & 1521 \\
Signature-based sampling & 27 & 640 \\
\hline
\end{tabular}
\vspace{-0.5em}
\end{table}

\section{Conclusions}
\label{sec:conclusion}
This work addressed the problem of fair cost allocation in ECs, where the DERs across multiple communities jointly influence DLMPs and overall dispatch outcomes. The key challenge was to ensure that operating cost savings from centralized scheduling are distributed fairly across communities, accounting for their contributions to system-level performance. To this end, a bilevel optimization framework was proposed from the perspectives of the CEA and the DSO. The bilevel problem was reformulated into a tractable single-level MILP using KKT conditions and strong duality. To ensure fair cost distribution among communities, the Shapley value was applied to quantify each community's marginal contribution to system-wide cost savings.

Simulation results demonstrated that the proposed Shapley-based allocation method fairly reflects the locational impact and resource contribution of each community. Furthermore, sensitivity studies showed that communities hosting larger BESS units receive higher cost savings, highlighting the incentive compatibility of the proposed approach.


\bibliographystyle{IEEEtran}
\bibliography{biblio.bib}

\end{document}